# Standard Model scales from warped extra dimensions

Bernard Riley

*AMEC Nuclear, 601 Faraday Street, Birchwood Park, Warrington WA3 6GN, UK*

**Abstract**

If in the Randall and Sundrum RS1 model the inverse of the compactification radius, the AdS curvature scale, and the five and four-dimensional Planck scales are equal in size, as is natural, then the warp factor at the location of the low energy brane is of value $1/\pi$. So that all scales derive from locations in the space, we identify the extra dimension with the infinite covering space of the $S^1/Z_2$ orbifold. The extra dimension is then essentially a series of connected line intervals, punctuated by branes. Scales on successive branes in the extra dimension descend from Planck scale in a geometric sequence of common ratio $1/\pi$. Evidence is provided for such a sequence within the spectrum of particle masses, and of a second geometric sequence, of common ratio $2/\pi$, which suggests that the AdS spacetime is six-dimensional and doubly warped. The scales of the Standard Model lie at coincident levels within the two sequences. A third sequence, of common ratio $1/e$, provides a symmetrical framework for the Standard Model and points to a warped product spacetime.

## 1. Introduction

There are many scales in our world. It seems natural that those scales should be related to the Planck scale through exponential factors. In the Randall and Sundrum RS1 and RS2 braneworld models [1, 2], our world, a 3-brane, is embedded in a five-dimensional spacetime. In RS1, there are two 3-branes, situated at the fixed points of an $S^1/Z_2$ orbifold. The two branes are separated by a slice of $AdS_5$ spacetime, of curvature $k$. The particles and forces of the Standard Model are confined to the Weak brane, while gravity is localised towards the Planck brane. The natural scale on the Weak brane is suppressed from Planck-scale by a warp factor $\exp(-ky)$, where $y$ is the coordinate of the extra dimension. In RS2, there is only one brane, the Planck brane, in an infinite extra dimension. Lykken and Randall have combined the RS1 and RS2 models so that gravity is localised on the Planck brane and our world is confined to a brane situated some distance away in an infinite extra dimension [3]. Scales in our world are exponentially suppressed from Planck scale. Oda has extended the Lykken and Randall model to include many branes upon which gravity is trapped [4]; our world is confined to one of those branes. In our model, we relate the scales of our world to locations in warped extra dimensions of infinite extent.

By hypothesising that particles occupy mass levels that descend from the Planck Mass in geometric sequence, and then computing the common ratio of the sequence, the hadrons have been found to occupy 'half-levels', 'quarter-levels', etc, within a sequence of common ratio $2/\pi$ [5]. The sublevels referred to above lie within geometric sequences of common ratio $(2/\pi)^{1/2}$, $(2/\pi)^{1/4}$, etc. Mass levels in geometric sequence is in some accord with a discretised version of RS1, in which the mass eigenstates of the lattice theory are geometrically spaced [6]. Eventually, a second geometric sequence of particle mass levels was identified, of common ratio $1/\pi$ [7].

In this letter, evidence is provided that particles of all kinds occupy mass levels within one or other of two geometric sequences that descend from the Planck Mass in geometric sequence. Three quark mass scales: u-d, s-c and b-t are defined and shown to lie at coincident levels and sublevels in the two sequences. The same sequence of scales includes the Higgs field VEV, the energy equivalent of the Rydberg constant, $R_\infty hc$ and the GUT scale of the MSSM. A third geometric sequence provides a symmetric framework for the Standard Model. We relate the three sequences to the geometry of a warped product spacetime $AdS_6 \times S^n$.

All values of particle mass have been taken from the 2008 Particle Listings of the Particle Data Group [8]. For the Planck Mass, we have used the current, 2006, CODATA recommended value, $1.220892 \times 10^{19}$ GeV. Particles are not distinguished from their antiparticles for ease of description of the phenomena. We refer to $\pi^\pm$, $K^\pm$, etc as single particles.

## 2. Particle mass levels

In figures 1 and 2, the following particles are shown to occupy levels, called principal levels, and half-levels within either Sequence 1, which descends from the Planck Mass with common ratio $1/\pi$, or Sequence 2, which descends from the Planck Mass with common ratio $2/\pi$: the charged leptons e, $\mu$ and $\tau$, the charged pseudoscalar mesons $\pi^\pm$, $K^\pm$ and $D^\pm$, the weak gauge bosons $W^\pm$ and $Z^0$, and the baryon singlet states $\Lambda$ (uds), $\Lambda_c^+$ (udc) and $\Lambda_b^0$ (udb).

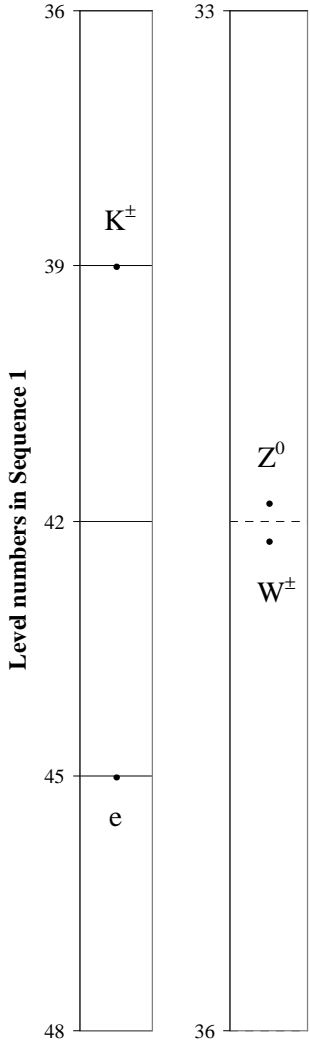

**Figure 1:** Occupation of Sequence 1

The electron and the $K^\pm$ meson occupy principal levels in Sequence 1. The muon and $\pi^\pm$ meson share a principal level in Sequence 2, as do the tau lepton and the $D^\pm$ meson; the particles in each of these apparent partnerships are arranged about the level. The weak gauge bosons share a half-level in Sequence 1. $\Lambda$ (I=0) shares a principal level in Sequence 2 with $\Sigma^0$ (uds, I=1). Again, the two particles are arranged about the shared level. $\Lambda_c^+$ (udc) and $\Lambda_b^0$ (udb) occupy half-levels in Sequence 2. Partnerships will be discussed in section 4.

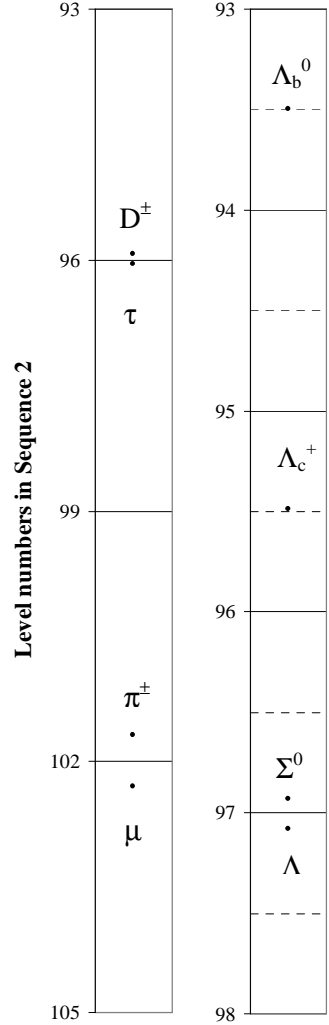

**Figure 2:** Occupation of Sequence 2

Particle mass levels may result from the geometry of spacetime. Consider the geometry of the RS1 model. A slice of $AdS_5$ spacetime is bounded by 3-branes at $y=1/k$ and $y=1/wk$, where $k$ is of order the Planck scale and $w$ is the warp factor. Assume that $1/r_c=k=M=M_P$ where $r_c$ is the compactification radius, $M$ is the five-dimensional Planck Mass and $M_P$ is the four-dimensional Planck Mass. This assumption is consistent with RS1, in which there are no large hierarchies. Since the 3-branes of the model lie at the fixed points of the $S^1/Z_2$ orbifold, our spacetime is bounded at $y=1/k$ and $y=\pi r_c=\pi/k$. And therefore $w=1/\pi$ at the location of the low energy 3-brane. The mass scale on the brane is $M_P/\pi$. We wish to construct a model with an infinite extra dimension so that four-dimensional scales derive from locations in the extra dimension. The extra dimension must therefore extend to $y > \pi r_c$, into the infinite covering space of the $S^1/Z_2$ orbifold. The extra dimension is then essentially a series of connected line intervals. With 3-branes punctuating the space at the boundaries of the line intervals, the geometry of the model resembles that of the multiple-branes model of Lykken and Randall [3], as developed by Oda [4]. With increasing $y$, the mass scales on successive 3-branes descend from the Planck Mass, the scale on the Planck brane at $y=1/k$, in a geometric sequence of common ratio $1/\pi$. This might be Sequence 1 for which evidence has been provided above. If so, it would seem that the matter fields of our world propagate on branes located in an extra dimension, at the boundaries referred to above and at subboundaries of obscure origin. Particles occupy the mass levels of two geometric sequences, Sequences 1 and 2, suggesting that the AdS bulk is six-dimensional and doubly warped. The two infinite extra dimensions of the model are the universal covers of the $S^1/Z_2$ and $S^1/Z_2 \times Z_2'$ orbifolds.

Some particle masses are associated with locations in one or other extra dimension that are identified in the base space.

### 3. The scales of the Standard Model

In figures 1 and 2, particles within partnerships were shown to be arranged about mass levels and sublevels. There is a tendency for the particles to take up a symmetric arrangement. If pairs of quarks are symmetrically arranged about levels or sublevels then the geometric mean of the two quark masses will coincide with a level or sublevel. The u-d scale is considered to be 5 MeV. The s-c and b-t scales have been calculated using the latest quark mass evaluations of the Particle Data Group [8]. These values have been used: $m_s$=95 MeV, $m_c$=1.25 GeV, $m_b$=4.2 GeV and $m_t$=172.5 GeV. In figure 3, the quark mass scales (5 MeV, 345 MeV and 26.9 GeV), the energy equivalent of the Rydberg constant, $R_\infty hc$ (13.6 eV) and the Higgs field VEV (246 GeV) are shown to lie at coincident levels and sublevels within the two sequences. There is a high degree of coincidence between Level 43 in Sequence 1 (5.120 MeV) and Level 109 in Sequence 2 (5.124 MeV). Level 28 in Sequence 1 ($1.47 \times 10^5$ GeV) and Level 71 in Sequence 2 ($1.45 \times 10^5$ GeV) meet closely, as do Level 58 in Sequence 1 (0.179 eV) and Level 147 in Sequence 2 (0.181 eV). Sublevels within the sequence of scales, which is of common ratio $(1/\pi)^{15}$, occur at coincident levels and sublevels in Sequence 1 and Sequence 2. For example, there is a half-level in the sequence of scales at Level 35.5 in Sequence 1 (27.4 GeV) and Level 90 in Sequence 2 (27.3 GeV). Another half-level within the sequence of scales has the value $2 \times 10^{16}$ GeV. This is the GUT scale in the MSSM.

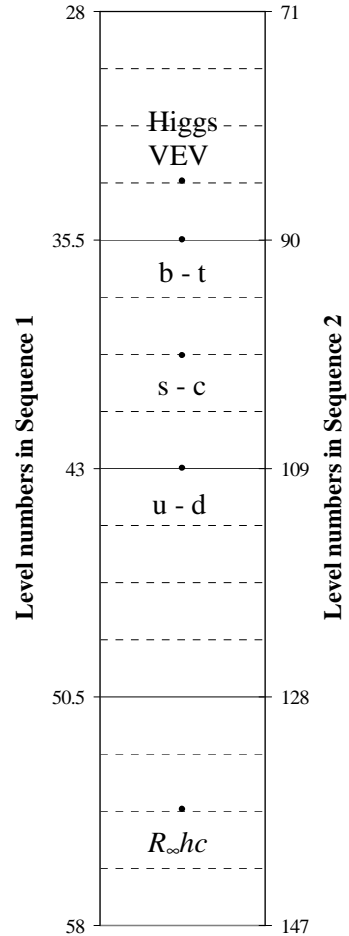

**Figure 3:** Scales of the Standard Model

### 4. Sequence 3

Sequence 3 descends from the Planck Mass with common ratio $e^{-1}$ and provides a symmetrical framework for the Standard Model, centred on Level 45, as shown in figure 4. Level 45 is of value 349 MeV, ≈ QCD scale. The neutron and $\pi^0$ meson are arranged symmetrically about Level 45. The Higgs field VEV and the electron mass stand in symmetric opposition either side of Level 45. The quark mass scales, u-d, s-c and b-t, defined in section 3, are also arranged symmetrically about Level 45. The s-c scale lies at Level 45. Level 50 in Sequence 3 has the value 2.35 MeV, which is the scale $M_L$ of the light matter content of the world. We define $M_L$ as the geometric mean of the masses of the up quark, down quark and electron, $(m_{ud}^2 m_e)^{1/3}$, where $m_{ud}$ is the u-d quark mass scale (5 MeV) and $m_e$ is the mass of the electron. Calculated in this way, $M_L$ has the value 2.34 MeV.

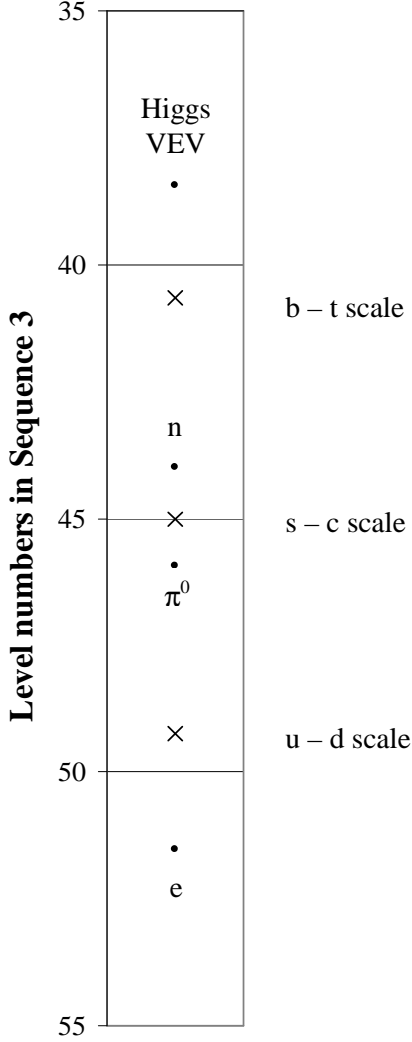

**Figure 4:** Sequence 3

The existence of Sequence 3 suggests that within spacetime there is another infinite warped space, with boundaries, as well as the universal covers of the $S^1/Z_2$ and $S^1/Z_2 \times Z_2'$ orbifolds. We conjecture that the base of this space is a warped n-sphere, reasoning simplistically that the warp factor on the sphere, radius $r$, has the value $e^{-kr} = e^{-1}$, where $k = 1/r$. Our spacetime has the geometry $AdS_6 \times S^n$. It might be the warped product $AdS_6 \times S^4$, which is a fibration of $AdS_6$ over $S^4$, and is the most general form of a metric that has the isometry of an $AdS_6$ spacetime [9, 10]. The warped product $AdS_6 \times S^4$ is a solution of massive ten-dimensional Type IIA string theory [11].

Finally, we return to the question of charged lepton – charged meson partnerships. In section 2, we saw that the muon and $\pi^\pm$ meson, and the tau lepton and $D^\pm$ meson appear to form partnerships centred upon principal levels in Sequence 2. Some strange hadrons, for example $\Lambda$ and $\Sigma^0$, also form partnerships centred upon levels or sublevels in Sequence 2 [5]. In these partnerships the mass difference of the two particles is precisely equal to the mass of a principal level in Sequence 2. Here though the mass differences of the muon and $\pi^\pm$ meson (33.91 MeV), the tau lepton and $D^\pm$ meson (92.5 MeV), and also the $\pi^0$ and $\pi^\pm$ mesons (4.594 MeV) are precisely equal to the masses of third-levels within Sequence 3, as shown in figure 5.

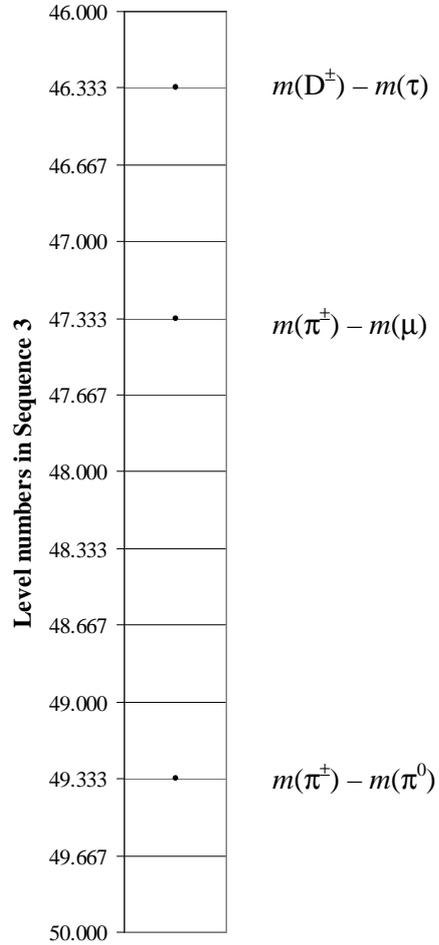

**Figure 5:** Particle mass differences

The three values of mass difference are related by factors of $e^n$, where $n$ is an integer, and are associated with locations in an extra space that are identified in the base of that space. The evidence suggests that the muon and $\pi^\pm$ meson, and the tau lepton and D$^\pm$ meson form partnerships characterised by $\Delta J = ½$.